\documentclass[12pt]{article}

\usepackage{graphicx}
\usepackage{epsf}

\usepackage{amsmath,amssymb,graphicx, float} 

\usepackage{pstricks}
\usepackage{color}

\def\beq{\begin{equation}}
\def\eq{\end{equation}}
\def\eeq{\end{equation}}

\def\centeron#1#2{{\setbox0=\hbox{#1}\setbox1=\hbox{#2}\ifdim
\wd1>\wd0\kern.5\wd1\kern-.5\wd0\fi
\copy0\kern-.5\wd0\kern-.5\wd1\copy1\ifdim\wd0>\wd1
\kern.5\wd0\kern-.5\wd1\fi}}
\def\ltap{\;\centeron{\raise.35ex\hbox{$<$}}{\lower.65ex\hbox{$\sim$}}\;}
\def\gtap{\;\centeron{\raise.35ex\hbox{$>$}}{\lower.65ex\hbox{$\sim$}}\;}
\def\gsim{\mathrel{\gtap}}
\def\lsim{\mathrel{\ltap}}

\def\chii0{\chi_i^0}
\def\chij0{\chi_j^0}

\def \V{\widetilde{V}}

\def\foursqr#1#2{{\vcenter{\vbox{
 \hrule height.#2pt
 \hbox{\vrule width.#2pt height#1pt \kern#1pt
 \vrule width.#2pt}
 \hrule height.#2pt
 \hrule height.#2pt
 \hbox{\vrule width.#2pt height#1pt \kern#1pt
 \vrule width.#2pt}
 \hrule height.#2pt
     \hrule height.#2pt
 \hbox{\vrule width.#2pt height#1pt \kern#1pt
 \vrule width.#2pt}
 \hrule height.#2pt
     \hrule height.#2pt
 \hbox{\vrule width.#2pt height#1pt \kern#1pt
 \vrule width.#2pt}
 \hrule height.#2pt}}}}
\def\psqr#1#2{{\vcenter{\vbox{\hrule height.#2pt
 \hbox{\vrule width.#2pt height#1pt \kern#1pt
 \vrule width.#2pt}
 \hrule height.#2pt \hrule height.#2pt
 \hbox{\vrule width.#2pt height#1pt \kern#1pt
 \vrule width.#2pt}
 \hrule height.#2pt}}}}
\def\sqr#1#2{{\vcenter{\vbox{\hrule height.#2pt
 \hbox{\vrule width.#2pt height#1pt \kern#1pt
 \vrule width.#2pt}
 \hrule height.#2pt}}}}

\def\figin{\epsfcheck\figin}\def\figins{\epsfcheck\figins}
\def\epsfcheck{\ifx\epsfbox\UnDeFiNeD
\message{(NO epsf.tex, FIGURES WILL BE IGNORED)}
\gdef\figin##1{\vskip2in}\gdef\figins##1{\hskip.5in}
\else\message{(FIGURES WILL BE INCLUDED)}%
\gdef\figin##1{##1}\gdef\figins##1{##1}\fi}
\def\DefWarn#1{}
\def\figinsert{\goodbreak\midinsert}
\def\ifig#1#2#3{\DefWarn#1\xdef#1{fig.~\the\figno}
\writedef{#1\leftbracket fig.\noexpand~\the\figno}%
\figinsert\figin{\centerline{#3}}\medskip\centerline{\vbox{\baselineskip12pt
\advance\hsize by -1truein\noindent\footnotefont{\bf
Fig.~\the\figno:\ } \it#2}}
\bigskip\endinsert\global\advance\figno by1}

\def\fig#1#2#3#4{\vskip 0.5cm \begingroup \midinsert \centerline{
\psfig{file=#1,width=#2}} \vskip 0.4cm
\global\advance\figno by 1
\centerline{\vbox{\baselineskip=12pt \noindent Figure \the\figno: #3}}
\endinsert \endgroup {\xdef#4{\the\figno}} }

\def\figcrop#1#2#3#4#5#6#7#8{\vskip 0.5cm \begingroup \midinsert \centerline{
\psfig{file=#1,width=#2,bbllx=#3,bblly=#4,bburx=#5,bbury=#6}} \vskip 0.4cm
\global\advance\figno by 1
\centerline{\vbox{\baselineskip=12pt \noindent Figure \the\figno: #7}}
\endinsert \endgroup {\xdef#8{\the\figno}} \vskip .5cm}

\def\figlabel#1{\xdef#1{\the\figno}}
\def\encadremath#1{\vbox{\hrule\hbox{\vrule\kern8pt\vbox{\kern8pt
\hbox{$\displaystyle #1$}\kern8pt}
\kern8pt\vrule}\hrule}}
\def\underarrow#1{\vbox{\ialign{##\crcr$\hfil\displaystyle
 {#1}\hfil$\crcr\noalign{\kern1pt\nointerlineskip}$\longrightarrow$\crcr}}}

\begin{document}

\begin{titlepage}

\begin{center}
\vspace*{-1cm}

\hfill RU-NHETC-2012-12 \\
\hfill UTTG-09-12 \\
%
\vskip 1.0in
{\LARGE \bf Jet Extinction from Non-Perturbative    } \\
\vspace{.25in}
{\LARGE \bf Quantum Gravity Effects  }\\
\vspace{.15in}

\vskip 0.65in
{\large Can Kilic$^1$},
{\large Amitabh Lath$^2$},
{\large Keith Rose$^2$},
{\large Scott Thomas$^2$}

\vskip 0.25in
$^1${\em
Theory Group, Department of Physics and Texas Cosmology Center\\
The University of Texas at Austin \\
Austin, TX 78712}

\vskip 0.15in

$^2${\em 
Department of Physics \\
Rutgers University \\
Piscataway, NJ 08854}

\vskip 0.75in
\end{center}

\baselineskip=16pt

\begin{abstract}
\noindent

The infrared-ultraviolet properties of quantum gravity suggest on
very general grounds that
hard short distance scattering processes are highly suppressed
for center of mass scattering energies beyond the fundamental Planck scale.
If this scale is not too far above the electroweak scale,
these non-perturbative quantum gravity effects could be manifest as an
extinction of high transverse momentum jets at the LHC.
To model these effects we implement an Extinction Monte Carlo
modification of the Pythia event generator based on a Veneziano
form factor with a large absorptive branch cut 
modification of hard QCD scattering processes.   
Using this we illustrate the leading effects of extinction on the
inclusive jet transverse momentum spectrum at the LHC.
We
estimate that an extinction mass scale of
up to roughly half the center of mass beam collision energy could
be probed with high statistics data.
Experimental searches at the LHC for jet extinction  
would be complementary to ongoing searches for the 
related phenomenon of excess production of high multiplicity final states.

\end{abstract}

\end{titlepage}

\baselineskip=17pt

\newpage





\section{Introduction}

The scattering of high energy particles in theories of quantum
gravity differs in qualitatively significant ways from scattering
in local quantum field theories
\cite{Banks:1999gd,Giddings:2011xs}.
The holographic infrared-ultraviolet properties of quantum gravity
imply that
at center of mass energies beyond the fundamental
Planck scale, scattering is dominated by non-perturbative
processes that result in final states with a high multiplicity of low energy particles.
In the very high energy limit, this
universal feature of quantum gravity scattering
can be understood in terms of semi-classical
formation of massive black hole intermediate states that subsequently
undergo low energy Hawking evaporation
\cite{Banks:1999gd}.
Concomitant with the predominance of high entropy non-perturbative states
at high energies is an exponential suppression of
perturbative hard scattering processes that give low multiplicity final states
\cite{Banks:1999gd,Giddings:2001bu}.
The essential extinction of hard 
short distance scattering processes in this regime
implies that
for incident colliding particles of any center of mass energy,
the energies of individual final state particles are effectively
bounded by the
fundamental Planck scale -- a striking
marker of the infrared-ultraviolet nature of quantum gravity.





If  the fundamental Planck scale is not too far above
the electroweak scale
\cite{ArkaniHamed:1998rs,Antoniadis:1998ig,Randall:1999ee}
certain
features of quantum gravity could in principle
be subject to experimental investigation at high energy colliders.
Most proposals for probing non-perturbative quantum gravity
effects in high energy collisions have focussed on
production of high entropy intermediate states, colloquially referred to
as microscopic black holes,
that decay to high multiplicity final states
\cite{Giddings:2001bu,Dimopoulos:2001hw,Meade:2007sz}.
At the Large Hadron Collider (LHC)
the CMS and ATLAS collaborations have conducted searches for
excess high multiplicity final states
\cite{Khachatryan:2010wx,Chatrchyan:2012taa,Aad:2011bw,Aad:2012ic}.
Other proposals have involved
long distance scattering in perturbative quantum gravity \cite{Giudice:1998ck,Mirabelli:1998rt}
and perturbative string theory effects
\cite{Cullen:2000ef,Lust:2008qc}.
The CMS and ATLAS collaborations have conducted searches for perturbative string excitations
and other purported quantum gravity resonance phenomena
in the di-jet mass
spectrum \cite{Khachatryan:2010jd,Chatrchyan:2011ns,ATLAS:2012pu,CMS:2012yf}.

In this paper we investigate the possibility of observing the extinction
of hard scattering processes in high energy collisions from non-perturbative
quantum gravity effects.
At the LHC the leading hard scattering process at high
energies is jet production from QCD interactions.
Extinction from non-perturbative quantum gravity effects
should therefore be most apparent as a suppression of
large invariant mass scattering processes
involving jets, or equivalently a suppression of
large transverse momentum
in the inclusive jet spectrum.
Such features should be a rather direct consequence of the
infrared-ultraviolet properties of non-perturbative quantum gravity
\cite{Banks:1999gd,Giddings:2001bu}.
Searches for the extinction of high energy hard scattering processes
would be complementary to searches for the appearance of high
multiplicity high energy processes.
Both features would be associated with scattering energies
far beyond the fundamental Planck scale.
Other features associated with physics near the fundamental scale,
such as resonance phenomena, depend on model dependent aspects
of the full theory of quantum gravity.
As a practical matter, current or planned high energy
colliders are unlikely to provide energies well in excess of the
fundamental Planck scale.
So it is worth exploring different signatures for the onset
of non-perturbative quantum gravity effects, since details
of which might provide the most sensitive probe at energies
of order the fundamental scale are likely to depend on specific attributes of the
full theory of quantum gravity.


The specific model we utilize to quantify the extinction of
hard scattering is a Veneziano form factor modification
of QCD processes with a large absorptive damping component
coming from a branch cut in the scattering amplitudes.  
This model can be motivated from some underlying features
of the Veneziano amplitude \cite{Veneziano:1968yb}.
The scattering of strings described by the Veneziano amplitude
is dominated at energies well above the string scale
by high entropy intermediate states \cite{Gross:1987kza,Gross:1987ar}.
For energies above the string scale, but below the Planck scale,
these states correspond to
excited perturbative strings, with
widths, or equivalently damping,
that grow with energy.
Above the Planck scale the excited states become strongly self coupled,
with large widths and damping.
In this regime the Veneziano amplitude
effectively represents scattering into a continuum of high entropy intermediate states
that have suppressed overlap with low multiplicity final states.
Imposing a large absorptive branch cut in the 
Veneziano amplitude essentially eliminates
the perturbative regime, and yields a model for the extinction
of low multiplicity scattering processes at energies above a fundamental scale.
This model is not meant to provide a complete description of
non-perturbative quantum gravity processes in specific theories
of quantum gravity, but only to provide a representation of
the onset of extinction in low multiplicity scattering processes.

One nice feature of the large damping Veneziano form factor
model employed here is that it smoothly matches
onto QCD as the fundamental scale is taken large.
This is in contrast to models with
black hole geometric cross sections
\cite{Giddings:2001bu,Dimopoulos:2001hw,Meade:2007sz}
that have been used to parameterize the results of LHC searches
for high multiplicity final states
\cite{Khachatryan:2010wx,Chatrchyan:2012taa,Aad:2011bw,Aad:2012ic}.
These models are well founded by general considerations
of semi-classical Einstein gravity for center of mass scattering energies
asymptotically beyond the fundamental scale.
But
the models do
not match smoothly onto lower energy scattering, and
their direct application to energies just above the
fundamental scale may be dubious.
In this sense quantifying the results of extinction searches
in terms of a fundamental scale
with the model utilized here could be considered to be more
conservative than those from
high multiplicity searches with the fundamental scale described
in terms of black
hole geometric cross sections.
It should also be noted that
the quantitative results from experimental searches for various types of
new physics associated with a
fundamental scale, such as
high multiplicity final states, resonance phenomena, and the extinction
effects presented here, can not generally be compared directly
since the models used to quantify these disparate processes make
different exclusive assumptions about what effects are important
near the fundamental scale.

The Veneziano form factor model for extinction of
leading order (LO) QCD scattering processes is detailed in the next section.
In section \ref{sec:LHC} we describe an Extinction Monte Carlo modification
of the Pythia event generator
based on the Veneziano form factor model.
We use this Monte Carlo to
simulate the leading effects of extinction on the
inclusive jet transverse momentum spectrum at the LHC,
and to roughly estimate the statistical sensitivity of 7 TeV and future
13 TeV measurements to the extinction mass scale.
We also suggest a simple and conservative
phenomenological form factor modification of next to leading order
(NLO) QCD predictions of the inclusive jet transverse momentum
spectrum
that could be used to quantify the results of
searches for extinction.


\section{Modeling Extinction}
\label{sec:modelextinction}

Simulation of jet extinction requires a phenomenological 
model for the modification of low multiplicity scattering processes.  
In the next subsection, form factor modifications 
for
all quark and gluon $2 \to 2$ scattering processes of QCD
based on a modified form of the Veneziano amplitude 
with large absorptive branch cut are introduced. 
These form factors 
provide a simple model representation 
of the general properties expected from the 
existence of strongly self coupled high entropy intermediate states.
In the following subsection  the salient quantitative
features of the extinction form factor model are presented.


\subsection{Veneziano Form Factor Model}
\label{sec:amps}

The $2 \to 2$ scattering amplitude for any QCD process
involving massless particles  $p_1,p_2,p_3,p_4$
may be represented
as a color ordered helicity amplitude ${\cal A}(p_1,p_2,p_3,p_4)_{\rm QCD}$.
The extinction model employed here is defined through
these scattering amplitudes modified by a universal 
form factor in adjacent color ordered kinematic
channels 
\beq
{\cal A}(p_1,p_2,p_3,p_4) =
{\cal A}(p_1,p_2,p_3,p_4)_{\rm QCD} ~\widetilde{V}(x_{12}, x_{23})
\label{amplitude-def}
\eq
where
$x_{AB} = (p_A  +  p_B)^2 = 2 p_A  \cdot  p_B$ with
all momenta incoming.
The purpose here is not to derive a complete expression for the 
form factor directly from some underlying theory of quantum gravity, 
but rather to develop a model expression based on general principles that captures 
the salient features of the onset of extinction.

In order to provide a useful model for the onset of extinction of 
$2 \to 2$ scattering processes,
the form factor $\widetilde{V}(x_{12}, x_{34})$ 
should satisfy certain general properties.  
The first is that the extinction effects should be characterized 
by an extinction mass scale $M_E$. 
And for the scattering amplitudes (\ref{amplitude-def}) to reduce to those of QCD 
at low energies, the form factor should approach unity 
for small values of the invariant momenta
\beq
\lim_{x_{12} ~ \! {\rm or}~ \! x_{34} \ll M_E} \widetilde{V}( x_{12},x_{34} ) = 1
\label{limreq}
\eq
The form factor may then be written in terms of a dimensionless function 
\beq
 \widetilde{V}( x_{12},x_{34} ) = V( x_{12}/M_E^2 ~ , ~ \! x_{34}/M_E^2 )
\eq
In order to represent the effects of extinction we also require 
that the form factor be bounded from above by unity 
\beq
V( x_{12},x_{34}) \leq 1
\label{extinctionprop}
\eq
with significant deviations from unity only for $x_{12},x_{34} \gsim 1$. 
A third requirement that we impose 
is crossing symmetry  
\beq
V(x_{12},x_{34} ) =ÊV(x_{34},x_{12} ) 
\label{crossing}
\eq
This ensures that extinction effects appear 
in all kinematic and color channels.  
As a simple ansatz satisfying the crossing 
symmetry property (\ref{crossing}) 
we consider form factors that factorize into a 
product of identical 
functions of 
the kinematic invariant in each  
channel, times an overall normalization 
that is a function of the sum of the kinematic invariants. 
The requirement (\ref{limreq}) of approach to unity for small values of 
the kinematic invariants then determines the form factor 
in terms of a single function 
\beq
V(x_{12} , x_{34} ) = 
{  V(x_{12})  V(x_{34}) \over 
  V(x_{12}  +  x_{34})    } 
  \label{dimlessff}
\eq
Another general requirement is dictated by the form of analytic continuation of the 
scattering amplitudes (\ref{amplitude-def}) for complex values of 
invariant momenta.  
The real parts of the QCD amplitudes are continuous 
when the kinematic invariants are extended
into the complex plane. 
Preserving this property for the modified amplitudes (\ref{amplitude-def})
requires that the form factor satisfies Hermitian analyticity 
\beq
V(z_{12}^*,z_{34}^* )  = 
\left[V(z_{12},z_{34} ) \right]^*
\label{HAcondition} 
\eq
This property introduces an absorptive branch cut in the 
imaginary part of the form factor, and is a crucial feature in modeling 
extinction of $2 \to 2$ scattering processes coming 
from the effects of 
high entropy intermediate states. 
Other requirements are provided by 
dispersion relations between the real and  
imaginary parts of the form factor.  
These integral relations constrain the behavior of the 
form factor for asymptotic values of the kinematic invariants.  
Since the purpose here is only to provide a phenomenological 
model for the onset of extinction that is applicable to 
kinematic invariants of order the extinction scale, $M_E$, 
we do not consider dispersion relation restrictions 
on the form factor that would apply
outside this kinematic regime.

\begin{figure}
\begin{center}
\includegraphics[scale=0.7]{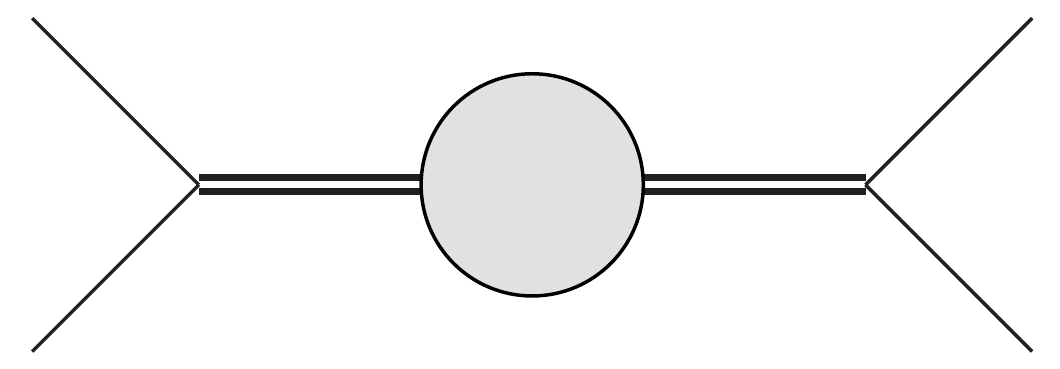}
\caption{Representation of the modification of QCD $2 \to 2$ scattering processes  
from the effects of heavy states that couple to intermediate multi-particle states. 
The single external lines represent quark and gluon scattering states.
The double lines represent heavy unstable states in the given kinematic 
and color channel.   
The grey blob represents high entropy intermediate states 
that produce a large absorptive branch cut in the 
imaginary part of the amplitude and lead to an 
extinction of the $2 \to 2$ scattering probability at high energies.  
}
\label{fig:blob}
\end{center}
\end{figure}

Local quantum field 
theory is not ameanable to a complete description 
of high energy scattering processes in quantum gravity. 
However, all the 
ingredients necessary for a phenomenological form factor model for the onset 
of extinction that satisfy the requirements listed above 
are available in local quantum field theory.  
So we employ this language to illustrate the elements of a model.  
We begin by introducing operators ${\cal O}_a$ that 
create and annihilate heavy unstable 
states with the same kinematic and color 
quantum numbers that appear in all channels of 
QCD $2 \to 2$ scattering processes.  
These operators couple to composite operators $J_{ai}$
that create and annihilate multi-particle states 
through interactions 
\beq
\sum_{a,i}~g_{ai} \int d^4x ~  J_{ai} {\cal O}_a~+~{\rm h.c.}
\label{heavyJO}
\eq
Among the operators $J_{ai}$ are ones
made out of quark and gluon fields that allow the heavy 
states to be exchanged in every kinematic and color channel 
in $2 \to 2$ quark and gluon scattering processes. 
A schematic representation of the modification of QCD 
$2 \to 2$ scattering processes from the 
intermediate multi-particle states through the 
exchange of the heavy states
is shown in Fig.~\ref{fig:blob}. 

The effects 
of the intermediate
multi-particle states in a given channel are contained within the 
normalized 
one-particle-irreducible 
two-point function for the composite operators 
that appear in that channel 
\beq
 i
\sum_{a,i}~ g^2_{ai} 
\int d^4x~e^{i p \cdot x} 
\langle 0 | T \{ J_{ai}(x) J_{ai}^{\dagger} (0) \} | 0 \rangle_{\rm 1PI} 
    \equiv 
 \left\{
    \begin{array}{rl} 
          - M^2(p^2) &  {\rm Boson} \\
          - M(p^2) &  {\rm Fermion} 
    \end{array} 
  \right.
  \label{twoptfcn}
\eq
In the space-like region $p^2 < 0$ and 
in the time-like region below threshold for production of intermediate 
states $p^2 < p_0^2 \geq 0$
there are no contributions from on-shell intermediate states, and 
the two-point function (\ref{twoptfcn}) is strictly real with no absorptive imaginary 
component.  
For the analytic continuation of the two-point function 
into the complex plane 
this implies that for ${\rm Re}(z)= p^2 < p_0^2$ along the 
real axis in this region 
\beq 
M^2 (z) = \left[ M^2(z^*) \right]^*
\label{analyticity} 
\eq
Extending the relation (\ref{analyticity}) to the entire 
complex plane implies that the 
imaginary component of the 
two-point function has a branch cut 
in the time-like region for real 
$x = p^2  \geq p_0^2$ on the real axis above threshold
of 
\beq
{\rm Disc}~M^2(x) \equiv 
M^2(x + i \epsilon) - M^2(x - i \epsilon) = 
2 i ~ {\rm Im}~M^2(x+i \epsilon) 
\label{branchcut}
\eq
The discontinuity is proportional to the total 
production rate $ \Gamma(p^2)$ of on-shell intermediate multi-particle states
from decay of the massive states
\beq
 \sqrt{ |p^2| }~ \Gamma(p^2) = {i \over 2} ~ {\rm Disc}~M^2(p^2) = 
- ~{\rm Im}~M^2(p^2+ i \epsilon)
\label{decayrate}
\eq
Unitarity requires that $i~{\rm Disc}~M^2(p^2 ) \geq 0$
along the branch cut.
The decay rate (\ref{decayrate}) 
represents a transfer  
of scattering probability from particular exclusive 
two-body final states to inclusive multi-particle 
final states.  
This feature 
allows for a representation of 
the effects of high entropy intermediate 
states in the phenomenological model through the absorptive 
branch cut (\ref{branchcut}).


%

The complete modification of QCD $2 \to 2$ scattering processes 
in a given kinematic channel 
from exchange of the 
heavy states with couplings (\ref{heavyJO}) to multi-particle 
intermediate states 
are 
obtained by summing the geometric series of all one-particle-reducible 
diagrams for the propagation of the heavy states in that channel 
connected 
by intermediate one-particle-irreducible multi-particle contributions to the 
two-point function (\ref{twoptfcn}). 
The $2 \to 2$ scattering diagram with two propagators for the heavy states 
connected by a single insertion of the two-point function 
(\ref{twoptfcn}) is illustrated in Fig.~\ref{fig:blob}.
The form factor function in a given kinematic channel 
arising from the geometric sum may be represented as a 
spectral decomposition over the heavy states
\beq
V(z ) =  1 + \int dm^2~
{  \rho(m^2) \over z - m^2 - M^2(z)}
\eq
where $\rho(m^2)$ is the spectral density 
of the heavy intermediate states with mass $m$. 
The real part of the two-point function (\ref{twoptfcn})
from the intermediate multi-particle states may be 
absorbed into the definition of the heavy state 
masses,  
 $m^{'2} = m^2 \! - \! {\rm Re}(M^2(z))$, 
so that the form factor has functional dependence 
$ V(z ) = f[ z \! -  i ~{\rm Im}(M^2(z))]$. 
Motivated by the general properties of the Venziano 
amplitude outlined in the introduction, 
we choose a gamma function with massless intercept 
as our model function to represent the effects of 
heavy intermediate states, 
$f(y) = \Gamma(1-y)$. 
With this model assumption, our 
dimensionless form factor (\ref{dimlessff}) is then 
\beq
V(z_{12},z_{34} ) =  { 
\Gamma \big[ 1 - z_{12} + i ~{\rm Im}(M^2(z_{12})) \big] 
\Gamma \big[ 1 - z_{34} + i ~{\rm Im}(M^2(z_{34})) \big] \over
 \Gamma \big[ 1 - z_{12} - z_{34} + i ~{\rm Im}(M^2(z_{12}+z_{34})) \big] 
 }
 \label{Vformimag}
\eq
The branch cut (\ref{branchcut}) 
in the two-point function (\ref{twoptfcn}) ensures that 
the form factor (\ref{Vformimag})
satisfies the Hermitian analyticity property (\ref{HAcondition})
for complex values of the kinematic invariants. 
For real values of the momenta, proper 
ingoing and outgoing boundary conditions 
are obtained from the form factor 
with kinematic invariants shifted just above the 
real axis, 
$V(x_{12},x_{34} ) \equiv V(x_{12} + i \epsilon ,x_{34} + i \epsilon )$ 
with $\epsilon \to 0$. 
For ${\rm Im}(M^2(z))=0$ the form factor (\ref{Vformimag})
reduces to the Veneziano amplitude \cite{Veneziano:1968yb}
with massless intercept. 

The final ingredient needed to specify the full functional 
dependence of the form factor is a 
model for the absorptive imaginary component of 
the two-point function (\ref{twoptfcn})
coming from intermediate on-shell multi-particle 
states. 
This absorptive component represents the transfer of 
scattering 
probability to inclusive multi-particle final states and is 
crucial in modeling the effects of extinction.  
Consistent with the analyticity condition (\ref{analyticity})
and multi-particle threshold at $x_0 = p_0^2 =0$ 
we parameterize the branch cut (\ref{branchcut}) along the positive real 
axis as 
\beq
- {i \over 2} ~ {\rm Disc}~M^2(x) = 
{\rm Im} ~M^2(x + i \epsilon) =  \alpha ~ x^n ~ \theta(x)  ~
\label{Imparam}
\eq
where $\theta(y)$ is the Heaviside step function. 
Following previous work \cite{Meade:2007sz} 
we adopt a phenomenological value of $n=1$ for the exponent 
of the momentum dependence of the branch cut.
Although this particular value for the exponent does not correspond  
to an analytic two-point function in the entire complex plane without 
additional unphysical branch cuts away from the real axis, 
it does provide a simple model parameterization for the  
magnitude of the absorptive phase in the physically 
important regime for extinction effects corresponding to  
kinematic invariants of order the extinction 
scale. 
With 
the parameterization (\ref{Imparam}) with $n=1$, the 
decay rate (\ref{decayrate}) of the massive states to on-shell 
multi-particle states is 
$\Gamma(m^2 ) = \alpha~m$.  
The damping parameter $\alpha$ therefore represents the dimensionless
width of the heavy states. 
The extinction property (\ref{extinctionprop}) is obtained for a 
constant $\alpha=1$. 
This value is used in the results and simulations presented below.

In the following subsections the squared amplitudes 
for all
QCD $2 \to 2$ scattering processes
organized by different choices of the external
particles with the form factor modification prescription 
(\ref{amplitude-def}) are presented.  
Similar results for the specific case of 
Veneziano form factor modifications 
without branch cuts 
may be obtained from leading order 
perturbative open-string scattering disk amplitudes 
\cite{Lust:2008qc}
but with some model dependence for amplitudes that involve 
only quarks and/or anti-quarks.   


\subsubsection{ Two Quarks of One Flavor and Two Quarks of Another Flavor}

The unique
color ordered helicity amplitudes with two quarks of one flavor
and two
quarks of another flavor may be written
\beq
{\cal A}(\bar{q}_1^{-i},q_2^{+j},\bar{q}_3^{\prime \pm k},q_4^{\prime \mp \ell}) =
A(\bar{q}_1^{-i},q_2^{+j},\bar{q}_3^{\prime \pm k},q_4^{\prime \mp \ell})
 ~(T^{a})_{j}^{~i} (T^{a})_{\ell}^{~k}
 \label{calAtwotwo}
 \eq
with others related by charge conjugation or parity, where throughout
helicity and momentum indicated in helicity amplitudes
are all incoming.
The $i,j,\dots$ are (anti)fundamental indices, $a,b,\dots$ are
adjoint indices, and $T^a$ are fundamental generators
of $SU(3)_C$ with
normalization
${\rm Tr}(T^a T^b )\! = \! \delta^{ab}$.
The QCD helicity amplitudes with color factors stripped off,
written in terms of twistor products are
related by permutations to a single amplitude
\beq
A(\bar{q}_1^{-i},q_2^{+j},\bar{q}_3^{\prime - k},q_4^{\prime + \ell})_{\rm QCD}  =
  g^2 ~
{  [ 13 ] \langle 42 \rangle \over
 \langle 12 \rangle  [ 21] }
 \label{twotwo-+-+}
\eq
\beq
A(\bar{q}_1^{-i},q_2^{+j},\bar{q}_3^{\prime + k},q_4^{\prime - \ell})_{\rm QCD}  =
  g^2 ~
{ [ 14 ] \langle 32 \rangle \over
 \langle 12 \rangle  [ 21] }
 \label{twotwo-++-}
\eq
where the twistor products are defined in terms of contractions of the wave function
spinors by
$[ AB ] \equiv \chi^{\alpha}(p_A) \chi_{\alpha}(p_B)$ and
$\langle AB \rangle \equiv \chi^*_{\dot{\alpha}}(p_A) \chi^{*\dot{\alpha}}(p_B) $
and satisfy
$\langle AB \rangle \! = \! -\langle BA \rangle $,
$\langle AB \rangle^* \! = \! [BA]$.
These color ordered helicity amplitudes may be averaged (summed)
over initial (final) state color and helicity configurations to give the
averaged squared matrix elements for $2 \to 2$ scattering processes
involving two quarks of one flavor and two of another flavor with
the form factor modification
\beq
\left\langle
|{\cal A}(q_1\bar{q}_2\rightarrow q'_3\bar{q}'_4) |^2
\right\rangle = g^4 ~
\frac{4}{9}\frac{u^{2}+t^{2}}{s^{2}} ~
|\V(s,t)|^2
\label{Mtwotwo-first}
\eq
\beq
\left\langle
|{\cal A}(q_1q'_2\rightarrow q_3q'_4) |^2
\right\rangle = g^4~
\frac{4}{9}\frac{s^{2}+u^{2}}{t^{2}} ~
|\V(t,u)|^2
\eq
\beq
\left\langle |{\cal A}(q_1\bar{q}'_2\rightarrow q_3\bar{q}'_4) |^2
\right\rangle = g^4 ~
\frac{4}{9}\frac{s^{2}+u^{2}}{t^{2}} ~
|\V(s,t)|^2
\eq
\beq
\left\langle |{\cal A}(\bar{q}_1\bar{q}'_2\rightarrow \bar{q}_3\bar{q}'_4) |^2
\right\rangle = g^4 ~
\frac{4}{9}\frac{s^{2}+u^{2}}{t^{2}} ~
|\V(t,u)|^2
\label{Mtwotwo-last}
\eq
where
$\langle AB \rangle [ BA ] \! = \! x_{AB}$, and
$s \! = \! x_{12} \! = \! x_{34}$,
$t \! = \! x_{13} \! = \! x_{24}$, and
$u \! = \! x_{14} \! = \! x_{23} $ are the
Mandelstam variables for the
scattering order $12 \to 34$ indicated
in the squared matrix elements.
Each scattering matrix element squared (\ref{Mtwotwo-first}-\ref{Mtwotwo-last})
includes a sum over two independent helicity channels (not related by parity) of
single helicity amplitudes squared
in a single color channel.


\subsubsection{Four Quarks of the Same Flavor}

The
color ordered helicity amplitudes with four quarks of the same flavor
with either all the chiralities the same or two of one chirality and
two of the other
may be written respectively
\beq
{\cal A}(\bar{q}_1^{-i},q_2^{+j},\bar{q}_3^{- k},q_4^{+ \ell}) =
A(\bar{q}_1^{-i},q_2^{+j},\bar{q}_3^{ - k},q_4^{+ \ell})
 ~(T^{a})_{j}^{~i} (T^{a})_{\ell}^{~k}
 \label{calAfoura}
\eq
\beq
{\cal A}(\bar{q}_1^{-i},q_2^{-j},\bar{q}_3^{+ k},q_4^{+ \ell}) =
A(\bar{q}_1^{-i},q_2^{-j},\bar{q}_3^{ + k},q_4^{ + \ell})
 ~(T^{a})_{j}^{~i} (T^{a})_{\ell}^{~k}
  \label{calAfourb}
\eq
with others related by charge conjugation or parity.
The QCD helicity amplitudes in this case
with the color factor stripped off
are identical to the amplitudes (\ref{twotwo-+-+}) and
(\ref{twotwo-++-}).
These color ordered helicity amplitudes
may be averaged (summed)
over initial (final) state color and helicity configurations to give the
averaged matrix elements squared for $2 \to 2$
scattering processes
involving four quarks of a single flavor with
the form factor modification
\beq
\left\langle
|{\cal A}(q_1\bar{q}_2\rightarrow q_3\bar{q}_4)|^2
\right\rangle
= g^4 ~
\frac{4}{9}\left[\frac{u^{2}+t^2}{s^{2}}+\frac{s^2+u^{2}}{t^{2}}-\frac{2}{3}\frac{u^{2}}{st}\right]
  ~
|\V(s,t)|^2
\label{Mfour-first}
\eq
\beq
\left\langle
|{\cal A}(q_1q_2\rightarrow q_3q_4)|^2
\right\rangle = g^4 ~
\frac{4}{9}\left[\frac{s^{2}+u^2}{t^{2}}+\frac{s^{2}+t^2}{u^{2}}-\frac{2}{3}\frac{s^{2}}{tu}\right]
    ~
|\V(t,u)|^2
\eq
\beq
\left\langle
|{\cal A}(\bar{q}_1\bar{q}_2\rightarrow \bar{q}_3\bar{q}_4)|^2
\right\rangle = g^4 ~
\frac{4}{9}\left[\frac{s^{2}+u^2}{t^{2}}+\frac{s^{2}+t^2}{u^{2}}-\frac{2}{3}\frac{s^{2}}{tu}\right]
    ~
|\V(t,u)|^2
\label{Mfour-last}
\eq
Each scattering matrix element squared (\ref{Mfour-first}-\ref{Mfour-last})
contains two types of terms.
The first type for two quarks of one chirality and two of the other
chirality includes a sum over two independent helicity channels (not
related by parity) of
single helicity amplitudes squared in a single color channel.
The second type for four quarks of the same chirality
includes a sum over the same two independent helicity channels
of the square of a coherent sum of two helicity amplitudes
in different color channels.



\subsubsection{ Four Gluons}

The unique four gluon
color ordered helicity amplitudes may be written
\beq
{\cal A}(g_1^{-a},g_2^{+b},g_3^{\pm c},g_4^{\mp d})
=
A(g_1^{-},g_2^{+},g_3^{\pm},g_4^{\mp})
~{\rm Tr}( T^a T^b T^c T^d)
\label{calAfourga}
\eq
\beq
{\cal A}(g_1^{-a},g_2^{-b},g_3^{+c},g_4^{+d})
=
A(g_1^{-},g_2^{-},g_3^{+},g_4^{+})
~{\rm Tr}( T^a T^b T^c T^d)
\label{calAfourgb}
\eq
with others related by parity.
The QCD helicity amplitudes with color factors stripped off are
all related by permutations to the Parke-Taylor amplitude \cite{Parke:1986gb}
\beq
A(g_1^{-},g_2^{+},g_3^{+},g_4^{-})_{\rm QCD} = g^2
\frac{\langle14\rangle^{4}}
{\langle12\rangle\langle23\rangle\langle34\rangle\langle41\rangle}
\eq
\beq
A(g_1^{-},g_2^{+},g_3^{-},g_4^{+})_{\rm QCD} = g^2
\frac{\langle13\rangle^{4}}
{\langle12\rangle\langle23\rangle\langle34\rangle\langle41\rangle}
\eq
\begin{equation}
A(g_1^{-},g_2^{-},g_3^{+},g_4^{+})_{\rm QCD} = g^2
\frac{\langle12\rangle^{4}}
{\langle12\rangle\langle23\rangle\langle34\rangle\langle41\rangle}
\end{equation}
These color ordered helicity amplitudes may be averaged (summed)
over initial (final) state color and helicity configurations to give the
averaged matrix elements squared for
$2 \to 2$ scattering processes
involving four gluons  with
the form factor modification
$$
\left\langle
|{\cal A}(g_1g_2\rightarrow g_3g_4)|^2
\right\rangle = g^4 ~
\frac{9}{4} \!  \left(\frac{1}{s^{2}}+\frac{1}{t^{2}}+\frac{1}{u^{2}}\right)
    \bigg[ s^{2}{|\V(t,u)|^{2}}+t^{2}{ |\V(s,u)|^{2}}+u^{2}
      { |\V(s,t)|^{2}}
$$
\beq
  ~~~~~~~~~~~~~~~~~~~~  ~~~~~~~~~~~~~~~~~~~~~~~~~~
  ~~~~~~~~~~~~
    - \frac{4}{27} \big| s~{ \V(t,u)}+t~{
         \V(s,u)}+u~{ \V(s,t)}\big|^{2} \bigg]
            \label{Mfourg}
 \eq
The scattering matrix element squared (\ref{Mfourg})
includes a sum over three independent helicity channels
(not related by parity) of
the square of a coherent sum of three helicity amplitudes
each in a different color channel.


\subsubsection{ Two Quarks and two Gluons}

The unique color ordered helicity amplitudes for
two quarks of a given flavor and chirality and two gluons may be written
\beq
{\cal A}(\bar{q}_1^{-i},q_2^{+j},g_3^{\pm a},g_4^{\mp b}) =
A(\bar{q}_1^{-},q_2^{+},g_3^{\pm},g_4^{\mp})
 ~(T^a T^b)_{j}^{~i}
 \label{calAtwotwog}
 \eq
with others related by charge conjugation or parity.
The QCD helicity amplitudes with color factors stripped off are
all related by permutations to a single amplitude
\beq
A(\bar{q}_1^{-},q_2^{+},g_3^{+},g_4^{-})_{\rm QCD} = g^2
\frac{  \langle14\rangle^{3}   \langle24\rangle   }
{\langle12\rangle\langle23\rangle\langle34\rangle\langle41\rangle}
\eq
\begin{equation}
A(\bar{q}_1^{-},q_2^{+},g_3^{-},g_4^{+})_{\rm QCD} = g^2
\frac{  \langle13\rangle^{3}   \langle23\rangle   }
{\langle12\rangle\langle23\rangle\langle34\rangle\langle41\rangle}
\end{equation}
These color ordered helicity amplitudes may be averaged (summed)
over initial (final) state color and helicity configurations to give the
averaged matrix elements squared for $2 \to 2$ scattering processes
involving two quarks of a given flavor and two gluons with
the form factor modification
\beq
\left\langle
|{\cal A}(q_1\bar{q}_2\rightarrow g_3g_4)|^2
\right\rangle \! = \! g^4 ~ \!
\frac{32}{27}\frac{u^{2}
  \! +\! t^{2}}{s^{2}}\left[\frac{u}{t}{
    |\V(s,t)|^{2}}
  \! + \!  \frac{t}{u}{ |\V(s,u)|^{2}}
  \! - \! \frac{1}{4}{
      {\rm Re}\big(\V(s,t)\V^{*}(s,u)\big)}\right]
      \label{Mtwotwog-first}
\eq
\beq
\left\langle
|{\cal A}(g_1g_2\rightarrow q_3\bar{q}_4)|^2
\right\rangle \! = \! g^4 ~ \!
\frac{1}{6}\frac{u^{2} \! + \! t^{2}}{s^{2}}\left[\frac{u}{t}{
    |\V(s,t)|^{2}} \! +\! \frac{t}{u}{ |\V(s,u)|^{2}} \! -\! \frac{1}{4}{
      {\rm Re}\big(\V(s,t)\V^{*}(s,u)\big)}\right]
\eq
\beq
\left\langle
|{\cal A}(q_1g_2\rightarrow q_3g_4)|^2
\right\rangle \! = \! g^4 ~ \!
\frac{4}{9}\frac{s^{2} \! +\! u^{2}}{t^{2}}\left[-\frac{s}{u}{
    |\V(t,u)|^{2}} \! -\! \frac{u}{s}{ |\V(s,t)|^{2}} \! +\! \frac{1}{4}{
      {\rm Re}\big(\V(s,t)\V^{*}(t,u)\big)}\right]
\eq
\beq
\left\langle
|{\cal A}(\bar{q}_1g_2\rightarrow \bar{q}_3g_4)|^2
\right\rangle \! = \! g^4 ~ \!
\frac{4}{9}\frac{s^{2} \! + \! u^{2}}{t^{2}}\left[-\frac{s}{u}{
    |\V(t,u)|^{2}} \! -\! \frac{u}{s}{ |\V(s,t)|^{2}} \! +\! \frac{1}{4}{
      {\rm Re}\big(\V(s,t)\V^{*}(t,u)\big)}\right]
            \label{Mtwotwog-last}
  \eq
Each scattering matrix element squared (\ref{Mtwotwog-first}-\ref{Mtwotwog-last})
includes a sum over two independent helicity channels
(not related by parity) of the square of a coherent sum
of two helicity amplitudes in different color channels.



\subsection{Behavior of the Form Factor}
\label{sec:ff}

The behavior of the Veneziano form factor model
defined in (\ref{Vformimag}) and (\ref{Imparam})
depends on the damping parameter $\alpha$.
For $\alpha < 1$ the form factor includes
Regge resonances
representing the effects of
intermediate perturbative string states, 
and does not satisfy the extinction property (\ref{extinctionprop}). 
For $\alpha = 1$ 
the extinction property (\ref{extinctionprop}) is obtained, with 
the modulus of the Veneziano factors
for central scattering
in all kinematic channels monotonically decreasing
as a function of the center of mass energy, 
whereas for $\alpha >1$ this feature is lost.
So for definiteness we take a universal value of $\alpha =1$
to model extinction of all QCD $2 \to 2$ scattering processes.
In this case the form factor with large absorptive branch cut 
effectively represents
a continuum of strongly self coupled high entropy intermediate states.
This is the regime that is relevant to use of the
Veneziano form factor as a model of extinction.

\begin{figure}
\begin{center}
\includegraphics[scale=0.5]{
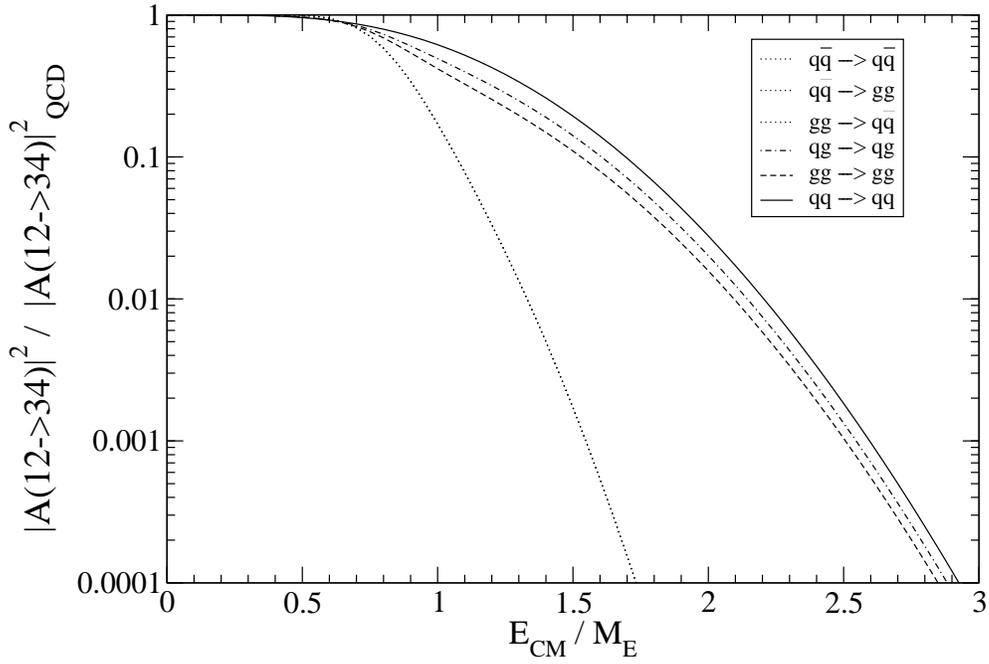}
\caption{Effects of
the Veneziano form factor extinction with $\alpha=1$
on various $2 \to 2$ quark and gluon scattering
probabilities
$|{\cal A}(12\rightarrow 34)|^2 / |{\cal A}(12\rightarrow 34)|^2_{\rm QCD}$
for central scattering $u \! =\! t  \! =\! -s/2$
as a function of the center of mass energy $E_{\rm CM} = \sqrt{s}$
in units of the extinction mass scale $M_E$.
}
\label{fig:ratio1D}
\end{center}
\end{figure}

The effects of the extinction Veneziano form factor modification of
various QCD $2 \to 2$
scattering probabilities with $\alpha =1$ for
central scattering corresponding to $t = u = -s/2$ are
shown in Fig.~\ref{fig:ratio1D}.
The rapid decrease
in effective overlap of the intermediate states with
low multiplicity states is apparent in the
extinction of central $2 \to 2$ scattering for
center of mass energies beyond the Veneziano extinction mass
scale $M_E$.
The extinction of scattering processes with both a quark
and anti-quark in the initial or final states is more
rapid than that for the other processes because the former
involve only $\widetilde{V}(s,t)$ or $\widetilde{V}(s,u)$,
while the latter include also $\widetilde{V}(t,u)$ which
falls more slowly for central scattering.

\begin{figure}
\begin{center}
\includegraphics[scale=0.75]{
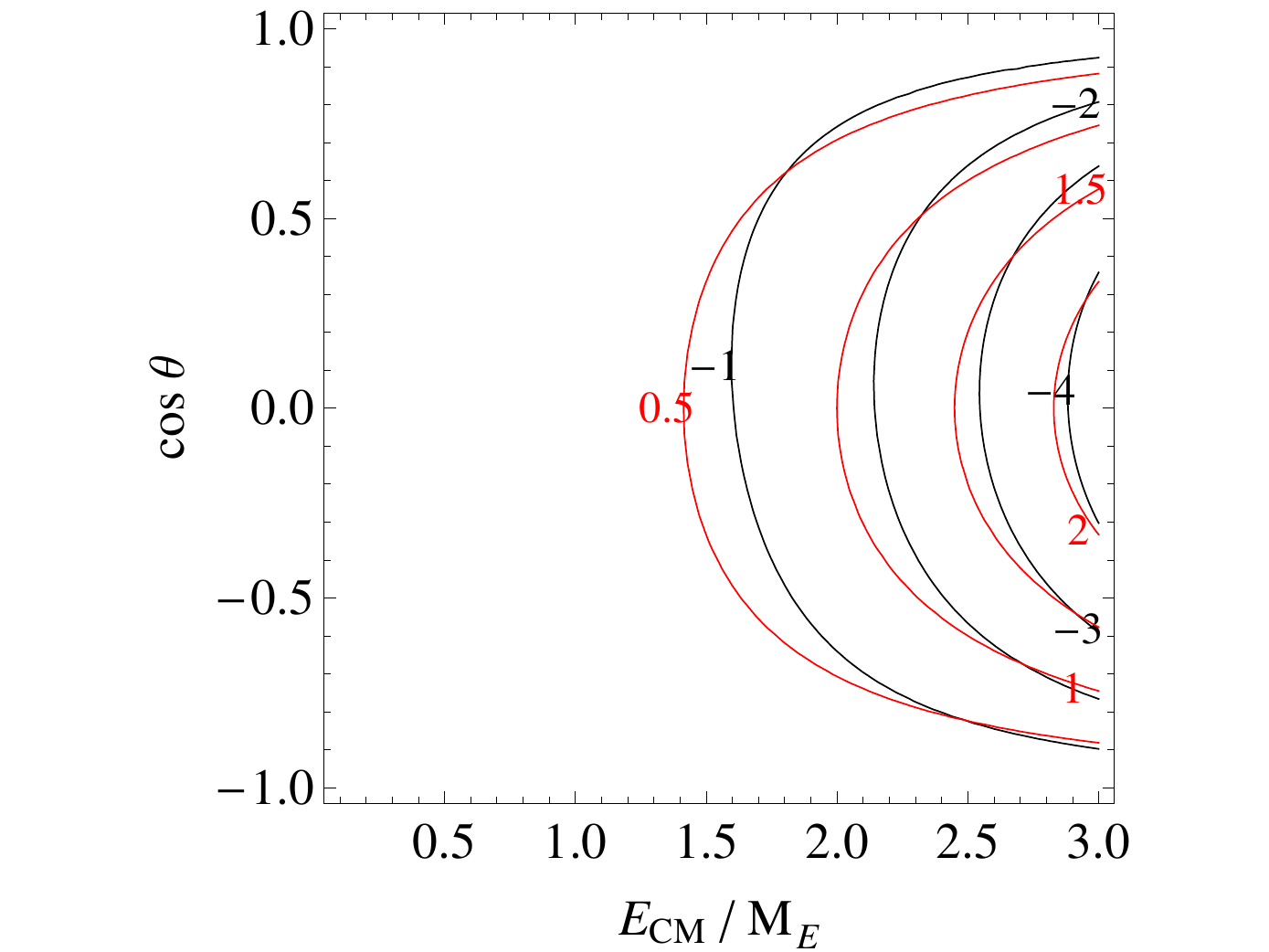}
\caption{
Effect of the Veneziano form factor extinction with $\alpha=1$ on the
$qg\rightarrow qg$ scattering probability as a function of the center of mass
mass energy, $E_{\rm CM} = \sqrt{s}$, and scattering angle,
$\cos \theta = 1 + 2t/s$.
Black contours indicate
$|{\cal A}(qg\rightarrow qg)|^2 / |{\cal A}(qg\rightarrow qg)|^2_{\rm QCD}$
labelled in $\log_{10}$ units.
Red contours indicate $p_T^2 / M_E^2 = (s/4)(1 - \cos^2 \theta) / M_E^2$
in linear units.
}
\label{fig:ratio2D}
\end{center}
\end{figure}

The dominant QCD $2 \to 2$ scattering process for proton-proton
collisions at high energy is $q g \to q g$.
The effects of the extinction Veneziano form factor modification
on this process with $\alpha=1$ are shown in Fig.~\ref{fig:ratio2D}
as a function of center of mass energy and scattering angle.
The extinction for scattering energies above the
Veneziano mass scale $M_E$ is most pronounced for
central scattering, with minimal extinction for
scattering near the forward and backward
directions.
The magnitude of extinction correlates well with the
transverse momentum of either of the scattered partons,
as shown in Fig.~\ref{fig:ratio2D}.
The transverse momentum is a rather direct measure of the
energy or length scale probed in the scattering
process.
And extinction
of hard scattering processes in proportion
to the transverse momentum
is expected on general grounds for scattering energies
above the fundamental Planck scale,
or equivalently scattering distances below the Planck length.
The fairly tight correlation between extinction and transverse
momentum is another nice feature of the Veneziano form
factor model of extinction used here.
And on general grounds, for scattering into final states of any multiplicity,
the transverse momentum
of the leading jet represents a useful kinematic
variable in which to characterize the degree of extinction.

\section{Jet Extinction at the LHC }
\label{sec:LHC}

Searching for jet extinction from non-perturbative quantum gravity
effects at the LHC requires a quantitative extinction model
to compare against data.
In the next subsection we describe an LO Extinction Monte Carlo
(EMC) modification of the Pythia Monte Carlo event generator
that implements
the QCD $2 \to 2$ scattering matrix
elements with Veneziano form factors presented in
section \ref{sec:modelextinction}.
In the following subsection we use the EMC to estimate
the statistical sensitivity of a search for extinction in the inclusive
jet transverse momentum spectrum.
We also describe how the results of the EMC could be
conservatively implemented as
a phenomenological extinction form factor
modification of NLO QCD predictions
for the inclusive jet transverse momentum spectrum.

\subsection{Monte Carlo Implementation}
\label{sec:MC}

In order to study quantitative effects of extinction at the LHC
with an EMC event generator, we incorporated
Veneziano form factors for QCD scattering processes
into Pythia~6.4~\cite{Sjostrand:2006za}.
The standard QCD $2 \to 2$ scattering processes in Pythia
appear in subroutine PYSGQC in the
form
\beq
{d \sigma \over d \hat{t} }(12 \to 34) =
{1 \over 16 \pi \hat{s}^2}
\left \langle |{\cal A}(12 \to 34)|^2 \right \rangle {1 \over S}
\eq
where $S=1(1/2)$ is a symmetry factor for distinct(identical) particles
in the final state, and $\hat{t}$ is integrated over the region
$- \hat{s} \leq \hat{t} \leq 0$.
We modified the squared matrix elements in
PYSGQC for each QCD $2 \to 2$ scattering process to include the
Veneziano form factors
given in section \ref{sec:amps}.
The gamma functions with complex arguments in the Veneziano amplitude were
evaluated numerically in a separate subroutine.
With this modification we are able to generate events corresponding
to a given extinction scenario with Veneziano form factor scale $M_E$
and damping parameter $\alpha$,
and interface the events to the (unmodified) parton
shower and hadronization subroutines of Pythia.
The leading initial and final state QCD radiation effects
in this extinction model
are therefore included in the Pythia event generation.
Since the energy scale for most radiation processes are well below
the hard scattering scale (which because of the rapid
extinction is effectively
at most of order the extinction scale)
this treatment of radiation effects should be a good
approximation, and amounts
formally to including next to leading logarithmic (NLL) effects
without extinction.
This realization of an EMC has the advantage that
in the limit of large extinction scale $M_E$,
the results reduce
to those of
Pythia 6.4, which has been well tested and validated.

To model detector effects, such as acceptance and jet
energy resolution, the hadronized events are passed
through the PGS detector simulation~\cite{PGS}.
We use the default
CMS parameter set of PGS, with a jet definition based on a cone algorithm with cone size $R =0.7$.


\subsection{Searching for Jet Extinction}

The general infrared-ultraviolet properties of non-perturbative
quantum gravity imply that
hard scattering processes should be suppressed for scattering
energies above the fundamental Planck scale.
Any observable used in a search for extinction at the LHC must be 
measured as a function of some kinematic invariant 
that characterizes the hard energy scale 
in the scattering process. 
The simplest and most inclusive kinematic variable that 
characterizes the energy scale in QCD scattering at a hadron collider 
is jet transverse momentum.  
The simplest inclusive observable as a function of this 
energy scale is the overall rate.  
And as discussed above, the extinction effects in the Veneziano form 
factor model with large absorptive branch cut are well 
correlated with transverse momentum.  
We therefore propose that a search for extinction could be
accomplished by a careful comparison of a high integrated luminosity
measurement of the
inclusive jet transverse momentum spectrum with
QCD expectations.
This spectrum is a robust inclusive observable
that can be calculated rather reliably in QCD at NLO
\cite{Nagy:2001fj,Nagy:2003tz,Kluge:2006xs}
and has already been measured at the LHC by the CMS
and ATLAS collaborations
with relatively low integrated luminosity
\cite{CMS:2011ab,Aad:2011fc}.
An extinction search
of this type could benefit from experience gained
in these previous measurements, and could be
carried out as an interpretation of
future high luminosity measurements of the inclusive
jet transverse momentum spectrum.

\begin{figure}
\begin{center}
\includegraphics[scale=0.5]{
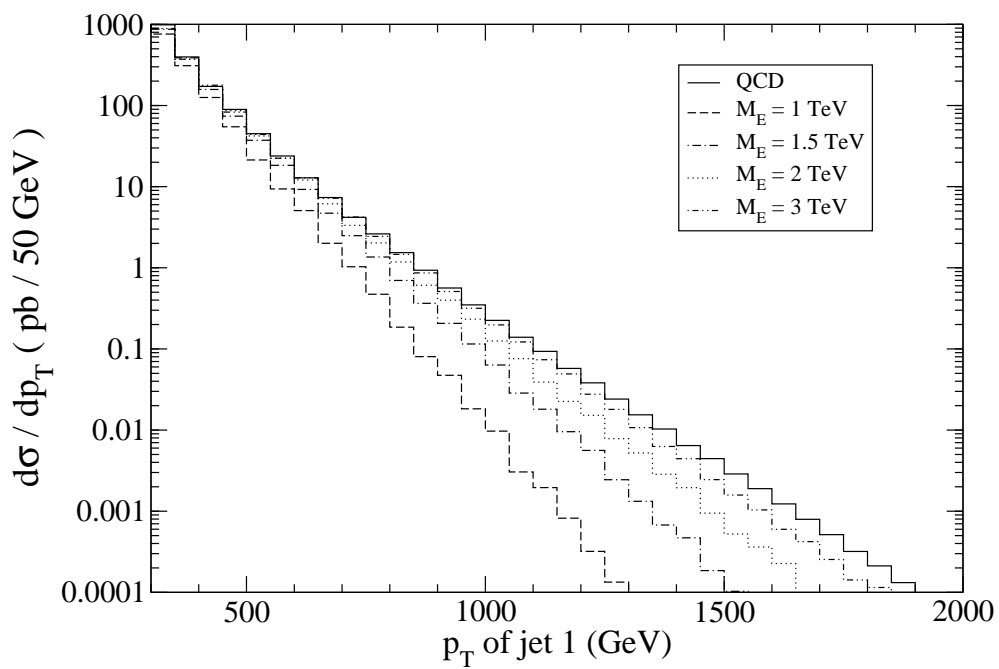}
\caption{The differential cross section
as a function of the transverse momentum of the leading jet for
various values of the Veneziano extinction mass scale $M_E$ with $\alpha=1$
for
7 TeV proton-proton collisions
after the selection cuts described in the text.}
\label{fig:EXT7TeV}
\end{center}
\end{figure}

\begin{figure}
\begin{center}
\includegraphics[scale=0.5]{
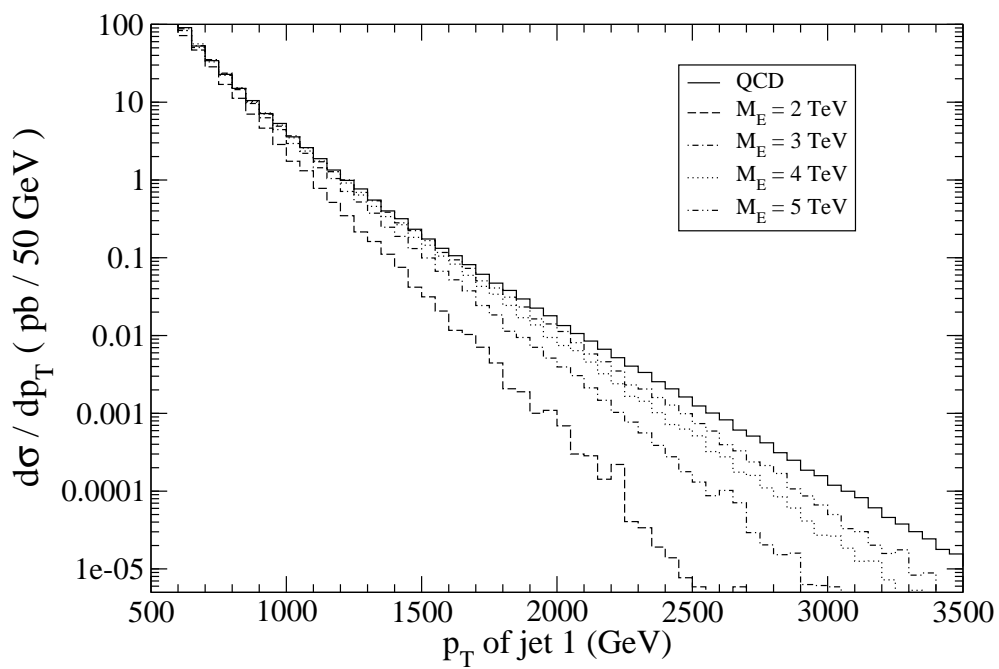}
\caption{
The differential cross section
as a function of the transverse momentum of the leading jet for
various values of the Veneziano extinction mass scale $M_E$ with $\alpha=1$
for
13 TeV proton-proton collisions
after the selection cuts described in the text.
}
\label{fig:EXT13TeV}
\end{center}
\end{figure}

In order to illustrate the leading effects of extinction at the LHC on the
inclusive jet transverse momentum spectrum we
use the EMC modification of the Pythia event generator
described in the previous subsection.
For event selection we require at least two jets, each with
transverse momentum $p_T > 500$ GeV and
pseudo-rapidity $|\eta| < 1.5$.
This yields hard scattering events with relatively large
center of mass scattering angle.
We simulate proton-proton collisions for both 7 TeV and 13 TeV
center of mass energies.
For 7 TeV we generate QCD events without any extinction
in $\sqrt{\hat{s}}$ bins of 300~GeV, starting at $\sqrt{\hat{s}}=200$~GeV,
each bin containing $50000$ events.
We use the same $\sqrt{\hat{s}}$ bins to simulate extinction with 25000 events in each bin for
Veneziano mass scales of $M_E=1,~1.5,~2$, and 3 TeV with $\alpha=1$.
For 13 TeV we use $\sqrt{\hat{s}}$ bins of 500~GeV
starting at $\sqrt{\hat{s}}=500$ GeV,
with $100000$ events in each bin for QCD, and $25000$ events per bin
for Veneziano mass scales of $M_E=2,~3,~4$, and 5 TeV with $\alpha=1$.
The differential cross section
spectra as a function of the leading jet transverse momentum
for these cases are shown in Figs. \ref{fig:EXT7TeV} and
\ref{fig:EXT13TeV}.
The extinction spectra drop to half of the QCD expectation
for a transverse mass of roughly one half of the extinction mass scale.

The statistical significance of the extinction of the
inclusive jet transverse momentum spectrum
as compared with QCD expectations
for
a given Veneziano extinction mass scale, center of
mass collision energy, and integrated luminosity
can be assessed with a simple likelihood analysis.
To do this we divide the leading jet transverse momenta
into 100 GeV bins above 500 GeV for 7 TeV collisions,
and above 1500 GeV for 13 TeV collisions.
Using the calculated values of the differential cross sections
we define probability distributions across the bins for QCD and the
extinction hypotheses labeled by the Veneziano mass scale $M_E$
\beq
P(i|M_E) = { d \sigma(i|M_E) / d p_T \over
  \sum_i d \sigma(i|M_E) / d p_T}
\eq
where $i$ labels the 100 GeV wide $p_T$ bin, and
where QCD corresponds to $M_E = \infty$.
Note that the overall normalization drops out, so
the probability distributions $P(i|M_E)$ are sensitive
to shape only.
A joint
likelihood entropy functional between a probability
distribution for a given extinction scale $M_E$
and a data set corresponding to an
extinction scale $M_E^\prime$ obtained with an
integrated luminosity $L$
may be defined
as being proportional to the logarithm
of the total joint probability
\beq
{\cal S}[ P(i|M_E) , N(i|M_E^\prime,L)   ]= - 2 \ln \Big( \prod_{i}  P(i|M_E)^{N(i|M_E^\prime,L)} \Big) =
- 2 \sum_i N(i|M_E^\prime,L) ~\ln P(i|M_E)
\eq
where $N(i|M_E^\prime,L)$
is the number of events in the $i$-th
$p_T$ bin.
For any data set corresponding to an extinction scale $M_E^\prime$,
the difference in likelihood entropy functionals for
probability distributions corresponding to extinction
scale $M_E$
and QCD, or equivalently the logarithmic likelihood ratio of extinction
to QCD,
\beq
Q[ P(i|M_E), P(i|{\rm QCD}) , N(i|M_E^\prime,L)   ] =
{\cal S}[ P(i|M_E) , N(i|M_E^\prime,L)   ] -
{\cal S}[ P(i|{\rm QCD}) , N(i|M_E^\prime,L)   ]
\label{Qeq}
\eq
may be used to
quantify the statistical separation between an extinction hypothesis
and QCD.

For each extinction mass scale
presented in Figs.~\ref{fig:EXT7TeV} and
\ref{fig:EXT13TeV} and for QCD
we generated normalized statistical probability distributions
\beq
{\cal P}(Q|M_E,L) \equiv { dP(Q|M_E,L) \over dQ }
\eq
of the joint likelihood entropy
differences
\beq
Q[N(i|M_E,L)] \equiv Q[ P(i|M_E), P(i|{\rm QCD}) , N(i|M_E,L)   ]
\eq
based on data from
50000 pseudo-experiments,
where the number of events $N(i|M_E,L)$ in each $p_T$ bin
of width $ \Delta p_T$
is chosen randomly from a
Poisson distribution with average expected events
$\langle N(i|M_E,L) \rangle =L \cdot \Delta p_T \cdot d \sigma(i|M_E) / d p_T $.
For 7 TeV collisions we take an integrated luminosity of
$L = 5$ fb$^{-1}$ and for 13 TeV collisions
$L = 100$ fb$^{-1}$.
Using a Gaussian fit, $\widetilde{\cal P}(Q|M_E,L)$, to each distribution we determine
the value of the likelihood entropy difference $Q_0(M_E,L)$
where the extinction and QCD data set distributions have
the same statistical probability for a given integrated luminosity
\beq
\widetilde{\cal P}(Q_0|M_E,L) = \widetilde{\cal P}(Q_0|{\rm QCD},L)
\eq
Data sets with
values of $Q[N(i|M_E,L)]>Q_0(M_E,L)$ are
more likely to be consistent with QCD, and ones with
$Q[N(i|M_E,L)]<Q_0(M_E,L)$ are more likely to be consistent with extinction.
The statistical probability that a QCD data set
is more consistent with an extinction hypothesis with extinction scale $M_E$
than with QCD is then
\beq
P({\rm QCD}|M_E,L) = \int_{-\infty}^{Q_0} dQ~\widetilde{\cal P}(Q|{\rm QCD},L)
\eq
and that an extinction data set with extinction scale $M_E$
is more consistent with the QCD hypothesis
than extinction is
\beq
P(M_E |{\rm QCD},L) = \int_{Q_0}^{+\infty} dQ~\widetilde{\cal P}(Q|M_E,L)
\eq
These two mis-identification probabilities need not be equal,
and to be conservative we consider the larger of the two.
For 7 TeV collisions with
5 fb$^{-1}$ of integrated luminosity we find that the statistical
mis-identification probability is below $5.7 \times 10^{-7}$
corresponding to $5 \sigma$ significance for an extinction
scale of $M_E \leq 3$ TeV.
And for 13 TeV
 collisions with
100 fb$^{-1}$ of integrated luminosity we find similar statistical
significance for $M_E \leq 5 $ TeV.
These rather conservative estimates
include statistical uncertainties only.
An
experimental search would also include detector systematic effects
such as from jet energy scale and resolution uncertainties that
are beyond the scope of this illustrative study.
But based on the results presented here,
it seems reasonable to expect that high luminosity
searches could probe extinction scales up to roughly
half the center of mass beam collision energy.

The Veneziano form factor model presented in section
\ref{sec:modelextinction} and utilized above
includes extinction effects only
for QCD $2 \to 2$ hard scattering processes.
Leading initial and final state QCD radiation effects
in this model
are included formally at NLL approximation
in the EMC modification of Pythia discussed
in section \ref{sec:MC}.
Full NLO effects for extinction of hard $2 \to 3$
and higher multiplicity scattering
processes are, however, not included.
Experimental measurements of the inclusive jet
transverse momentum spectrum \cite{CMS:2011ab,Aad:2011fc}
do make comparisons with available NLO QCD calculations
\cite{Nagy:2001fj,Nagy:2003tz,Kluge:2006xs}.
So it would be useful to present the effects of extinction
as a modification of these NLO predictions
of the form
\beq
{d \sigma  \over dp_T} (pp \to {\rm jets}~|~ p_T,M_E) \simeq
F(p_T,M_E) ~ {d \sigma \over dp_T}
(pp \to {\rm jets}~|~ p_T)_{\rm NLO}
\eq
where $F(p_T,M_E)$ is a phenomenological
form factor function that depends on the extinction scale $M_E$.
As a practical matter the magnitude of the
extinction scale that can
be probed in proton-proton collisions
is limited roughly by the highest transverse momenta
jets
where the data begins to run out.
In this regime where scattering energies are by definition
at most of order
the extinction scale, the leading effects of extinction
on energetic grounds are for the leading
$2 \to 2$ hard scattering processes.
Higher multiplicity processes at the
same transverse momentum
probe lower individual scattering energies
with smaller extinction effects.
So a conservative ansatz that would be useful
in quantifying the onset of extinction
would be to assume that extinction is limited
to $2 \to 2$ hard scattering processes.
Inclusion of extinction effects for higher multiplicity hard
scattering processes would only soften the
transverse momentum spectrum.
For a given data
set that is roughly consistent with NLO QCD expectations,
inclusion of these sub-dominant effects would only yield
a stronger bound.
This ansatz amounts to an $F(p_T,M_E)$ of the form
\beq
F(p_T,M_E) = \big[1 - C_{\rm NLO}(p_T) \big]
  f_{\rm LO}(p_T,M_E)~+~C_{\rm NLO}(p_T)
\eq
where $f_{\rm LO}(p_T,M_E)$ represents extinction of LO $2 \to 2$
scattering procesess, and
the $p_T$ dependent NLO fraction for $pp \to {\rm jets}$ is
\beq
C_{\rm NLO}(p_T) = 1 -
  {  (d \sigma / dp_T )_{\rm LO}  \over
     (d \sigma / dp_T )_{\rm NLO}   }
\eq
NLO QCD calculations of the inclusive jet transverse momentum
spectrum \cite{Nagy:2001fj,Nagy:2003tz,Kluge:2006xs}
generally include a numerical comparison of LO and NLO results
that may be used to extract  $C_{\rm NLO}(p_T)$.
A functional form for the leading order extinction
function $f_{\rm LO}(p_T,M_E)$ may be obtained by comparing
the
results of the EMC modification of Pythia described in section
\ref{sec:MC} for different $M_E$ with the Pythia QCD results for $M_E=\infty$.
The modulus squared of the Veneziano form factor
for central scattering in all kinematic channels has the general form of
a sigmoid function.
We have found that a simple sigmoid
gives a good functional fit to the
$p_T \lsim {\cal O}(M_E)$ onset of LO extinction
in the Veneziano model
\beq
f_{\rm LO}(p_T,M_E) = {1 \over
1 + {\rm exp} \big[ (p_T - p_{T,1/2} ) / p_{T,0} \big] }
\eq
where $p_{T,1/2} = f_{1/2}(M_E)$ and
$p_{T,0} = f_0(M_E)$ are $M_E$ dependent functions
that must be determined by fitting to Monte Carlo results.
The function $f_{1/2}(M_E)$ is the value of $p_T$ for a given
$M_E$ where
the LO inclusive jet spectrum drops to half of the QCD expectation,
and should be linear in $M_E$.
The function $f_0(M_E)$ is the rate of change of the extinction with $p_T$,
and should be well fit by constant plus linear term in $M_E$.
This phenomenological form factor modification of
NLO QCD predictions should
be sufficient to quantify the results of
future searches for the onset of extinction in the
inclusive jet transverse momentum spectrum.


\section{Conclusions}

The infrared-ultraviolet holographic properties of quantum
gravity suggest on very general grounds that
at scattering energies beyond the fundamental Planck scale,
low multiplicity
final states suffer rapid extinction.
If the fundamental Planck scale is not too far above the electroweak scale,
this interesting feature of non-perturbative
quantum gravity could in principle be subject to
experimental investigation in the laboratory.
In this paper
we proposed that these extinction effects could be
searched for in the inclusive
jet transverse momentum spectrum at the LHC.
We presented a large damping Veneziano form factor
extinction model
of QCD $2\to2$ scattering processes,
and implemented this
into a modified version of the Pythia event generator to simulate
extinction processes.
This model has the nice feature that it smoothly reduces
to Pythia QCD scattering as the Veneziano extinction scale
is taken large.
Based on simulations using these tools, we estimate that
future high luminosity measurements
of the inclusive jet transverse momentum spectrum
could be statistically sensitive to an extinction scale of order half of the
beam center of mass collision energy.
We also suggested a simple and
conservative phenomenological form factor modification of NLO
QCD predictions for this spectrum that could be used
to quantify future searches for the onset of extinction.

The search for jet extinction from non-perturbative
quantum gravity effects in high energy collisions at the LHC
is complementary to ongoing searches for the
related phenomenon of excess production of high multiplicity
final states.
Since the intrinsic sensitivity of these searches
depends on details of the underlying
theory of quantum gravity, both should be
part of the search program for
strong gravitational effects at the LHC.
Observation of either, or both, would open up a
fantastic new experimental window into the holographic
properties of gravity.


\bigskip
\bigskip

\noindent
{ \Large \bf Acknowledgments}

\smallskip \smallskip

\noindent
We would like to thank J.P. Chou for useful comments.
This research was supported in part by DOE grant
DE-FG02-96ER40959 and NSF Grant PHY-0969020.



\end{document}